\documentclass[prd,aps,preprint,nofootinbib,showpacs,floatfix]{revtex4}
\usepackage{graphicx,amssymb,amsmath}
\usepackage{epsf}
\usepackage{pslatex}
\usepackage{dcolumn}
\def\mpi2{m_\pi^2}
\def\mK2{m_K^2}

\newcommand{\bea}{\begin{eqnarray}}
\newcommand{\eea}{\end{eqnarray}}
\newcommand{\be}{\begin{equation}}
\newcommand{\ee}{\end{equation}}

\newcommand{\VEV}[1]{\left\langle #1\right\rangle}

\newcommand{\Tr}{\mbox{Tr}}

\newsavebox{\DERIVBOXZLM}
\savebox{\DERIVBOXZLM}[2.5em]{$\Longrightarrow\hspace{-1.5em}
\raisebox{.2ex}{*}
\hspace{-.7em}\raisebox{-.8ex}{\scriptsize lm}\hspace{.7em}$}

\begin{document}
\bibliographystyle{apsrev}
\epsfclipon

%%%%%%%%%%%%%%%%%%%%%%%%%%%%%% MACROS %%%%%%%%%%%%%%%%%%%%%%%%%%

\newcommand{\pbp}{\langle \bar \psi \psi \rangle}
\newcommand{\pbdmdup}{\left\langle \bar \psi \frac{dM}{du_0} \psi
\right\rangle}

%%%%%%%%%%%%%%%%%%%%%%%%%%%%%% TITLEPAGE %%%%%%%%%%%%%%%%%%%%%%%%%%

% \draft

\title{Charm annihilation effects on the hyperfine splitting in charmonium}

\author{L. Levkova$^1$ and C. DeTar$^1$} \affiliation{$^1$Physics Department, University of Utah,
Salt Lake City, Utah 84112, USA}

\begin{abstract}
In calculations of the hyperfine splitting in charmonium, 
the contributions of the disconnected diagrams
are considered small and are typically ignored. We aim to estimate nonperturbatively the size of
the resulting correction,
which may eventually be needed in high precision calculations of the charmonium
spectrum. We study this problem  
in the quenched and unquenched QCD  
cases. On dynamical ensembles the disconnected charmonium propagators 
contain light modes which complicate the extraction of the signal at large distances.
In the fully quenched case, where there are no such light modes, the interpretation
of the signal is simplified. We present results from lattices with $a\approx 0.09$ fm 
and $a\approx 0.06$ fm. 
\end{abstract}
\maketitle

\newpage

%%%%%%%%%%%%%%%%%%%%%%%%%%%%%% INTRODUCTION %%%%%%%%%%%%%%%%%%%%%%%%%%

\section{Introduction}

Lattice calculations of the hyperfine splitting in charmonium usually 
ignore the contributions of the annihilation (disconnected)
diagrams to both the vector $J/\psi$
and the pseudoscalar $\eta_c$ states. This  simplification 
leads to an error, and our goal is to determine 
the actual value of the contributions. Perturbatively, the
contribution of these diagrams  in charmonium is expected to be small
due to the Okubo-Zweig-Iizuka suppression, especially for the vector state \cite{OZI}.
However, nonperturbative effects, such as the $U_A(1)$ anomaly \cite{u1}
and mixing with glueball and light hadronic states,
 might enhance it
enough so that it becomes a nonnegligible fraction of the hyperfine splitting. 
Previous calculations \cite{fs1,fs2} using two-flavor gauge ensembles very 
roughly estimated
the contribution to be within $\pm20$ MeV. They both confirm that there are 
significant difficulties in obtaining a signal for the disconnected diagrams
due to noise, especially for heavy quarks. 

In our work, the charm quarks are simulated 
with the clover fermion action with $\kappa_c$ tuned to
the physical charm quark mass.  The disconnected diagrams are calculated stochastically
 with spin- and color-diluted sources.
Our calculation improves on the previous ones in a number of ways. First, we use larger lattice volumes
($28^3\times96$, $48^3\times144$, and $40^3\times96$) and point-to-point (PTP) propagators,
which significantly improve our statistics
and signal-to-noise ratio. Point-to-point propagators
reduce the relative standard error over time-slice-to-time-slice (TTT) propagators by one to three orders of magnitude. 
Second, our gauge ensembles have 
much finer lattice spacings $a$. We work with lattices with $a\approx0.09$ fm (fine ensembles) 
and $a\approx 0.06$ fm 
(superfine ensemble). Table~\ref{tab:lat} gives the parameters of the ensembles.
And finally, we employ the unbiased subtraction technique \cite{unb}
in the stochastic estimators used to determine the disconnected correlators.
The success of this technique depends on the fast convergence
of the hopping parameter expansion of the clover Dirac operator used in the subtraction.
Considering that $\kappa_c$ is still small for the charm quark at these lattice spacings, 
we use
the terms of the expansion only up to third order in $\kappa_c$, which reduces the
standard deviation of the disconnected correlator by an
additional factor of  about four.

In this study we attempt to determine the size of the effects of the disconnected diagrams
on the mass of the $\eta_c$ only. Our previous studies \cite{mine,mine1} and our current work
 show that
the effect of the charm annihilation on the vector state are much smaller than 1 MeV; 
thus we ignore it here
and equate the hyperfine effects with the effects in the pseudoscalar only. 
Our calculations are done on two fully quenched and one dynamical ensemble with two light degenerate quarks 
and one
strange quark (2+1 dynamical flavors) in the asqtad formulation \cite{Orginos:1998ue}.
In the fully quenched case,
the disconnected $\eta_c$ correlator can have at most additional contributions from the $U_A(1)$ anomaly and 
close-lying glueball states. In the 2+1 flavor dynamical case, the disconnected correlator can also couple to
light hadronic states, which complicates the task at hand significantly. In both the 2+1 flavor dynamical 
and fully quenched cases
we ignore contributions to the disconnected correlators from sea charm quark loops . 
To the extent that the disconnected contribution is small (first order),
the sea charm quark effects are second order and so negligible at our
level of precision.  Our result that the contribution is, indeed, small
makes the calculation self consistent.

This paper is organized as follows. Section~II outlines the analytic 
framework in which we interpret our lattice data. Section~III discusses some
general properties of the disconnected propagators which follow from
our analyses. In Sec.~IV we give our fitting method for the disconnected propagator.
Section~V is dedicated to the specifics of the tuning of the charm quark mass.
The final section, VI, contains our results and conclusions.
\begin{table}[t]
\begin{tabular}{llcccc}
\hline\hline
Ensemble & $a$ [fm] & $m_l/m_s$ & Volume & $\kappa_c$& \# config. \\
\hline
QF & $\approx0.085$ & $\cdots$ &$28^3\times96$ & {\bf0.120}, 0.127& 410  \\
QSF& $\approx0.063$&$\cdots$ & $48^3\times144$ & {\bf0.125}, 0.130 & 415 \\
DF& $\approx0.086$ & $0.0031/0.031$& $40^3\times96$ & {\bf0.125}, 0.127& 766 \\
\hline\hline
\end{tabular}
\caption{Run parameters of the quenched fine (QF), quenched superfine (QSF) and dynamical fine (DF) ensembles
are shown. 
The bold values of $\kappa_c$ are the ones
obtained by tuning the $\eta_c$ mass and are used in this study. The nonbold $\kappa_c$
values are from our previous studies \cite{mine,mine1} and are listed for comparison.}
\label{tab:lat}
\end{table}

\section{General analytic framework}

\begin{figure}[t]
\begin{center}
\includegraphics[width=0.85\textwidth]{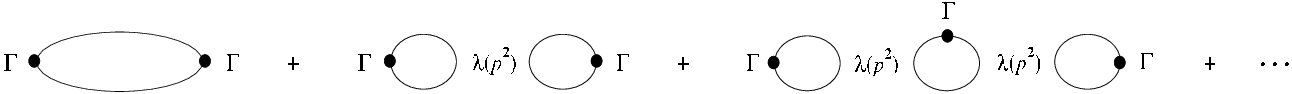}
\caption{Connected and disconnected diagrams contributing 
to the full propagator on lattices unquenched with respect to the charm
quark are shown.}
\label{fig:diagram}
\end{center}
\end{figure}

In this section we derive the shift of the mass of a flavor singlet state 
due to the contribution of the disconnected diagrams to its full propagator.
Figure~\ref{fig:diagram}
shows the diagrammatic expansion 
of the full propagator, where with continuous lines we represent the
charm quark propagators. 
The first term in this expansion is the connected piece
and the rest are disconnected propagators containing charm quark loops. 
We denote the momentum-space connected propagator of a (pseudo)scalar meson as
\be
C(p^2)=\frac{A}{p^2+m_c^2}\,,
\label{connprop}
\ee
where $A$ is a constant, and $m_c$
is the "connected" mass of the meson. (The vector meson propagator
has the same form as Eq~(\ref{connprop}), if we neglect the spin degrees of freedom).
Then the full propagator is the infinite sum:
\be
F(p^2)=\frac{A}{p^2+m_c^2} + \frac{\sqrt{A}}{p^2+m_c^2}\lambda(p^2)\frac{\sqrt{A}}{p^2+m_c^2} 
+\frac{\sqrt{A}}{p^2+m_c^2}\lambda(p^2)\frac{1}{p^2+m_c^2}\lambda(p^2)\frac{\sqrt{A}}{p^2+m_c^2}+\cdots\,,
\label{fullprop}
\ee
where the first term is the connected piece $C(p^2)$ and the rest are terms
which consist of disconnected quark loops of the same flavor as the meson's constituent quarks
(see Fig.~\ref{fig:diagram} for the diagrammatic representation of the three explicitly given terms).
The function $\lambda(p^2)$ effectively describes all possible interactions
between the quark loops in the disconnected pieces and the gauge fields, quarks of other flavors
or effects such as the $U_A(1)$ anomaly, if relevant for the specific meson state.
The disconnected propagator $D(p^2)$ is naturally the sum of all terms in Eq.~(\ref{fullprop}),
except the first one. 
After we sum the geometric progression in Eq.~(\ref{fullprop}), we obtain:
\be
F(p^2) = \frac{A}{p^2 +m_c^2 -\lambda(p^2)}= \frac{A}{p^2 +m_f^2}\,,
\ee  
where $m_f$ is the "full" meson mass we could calculate if we knew all 
terms that contribute to the full propagator. Thus, the difference between
the mass that is usually computed from only the connected propagator
and the actual mass is approximately
\footnote{The momentum dependence in $\lambda(p^2)$ leads to a first-order
  (in $\lambda$) renormalization of the connected pole residue, so there is
  a second order correction in our result that we safely ignore.}
\be
\delta m = m_c - m_f \approx \frac{\lambda(-m_c^2)}{2m_c}\,.
\label{dm}
\ee
In the last expression for simplicity we replaced the function $\lambda(p^2)$ with
the value of its largest contribution in Eq.~(\ref{fullprop}) at the pole $p^2=-m_c^2$.
Hence the sign of $\delta m$ 
depends on the sign of $\lambda(-m_c^2)$.
The mass shift $\delta m$ due to the disconnected quark loops can be treated as a
perturbation, in which case, to first order, both the connected and disconnected contributions can be
computed without dynamical sea quarks of the constituent's flavor (heavy-quark-quenched case).
In this case, only the
first two terms in Eq.~(\ref{fullprop}) survive and the disconnected propagator 
is reduced to the second term only
(shown diagrammatically in Fig.~\ref{fig:diagram}, middle).

\section{Properties of the disconnected propagator} 

The asymptotic behavior at large times $t$ of the full charmonium propagator, $F(t)$, is 
\be
F(t) = C(t) + D(t) = \sum_n\langle \Omega|O|n \rangle\langle n|O|\Omega \rangle e^{-E_nt}
\xrightarrow[t\rightarrow\infty]{} \langle \Omega|O|0 \rangle^2 e^{-E_0t}\,,
\ee
where the sum is over all eigenstates $|n\rangle$
of the Hamiltonian with corresponding energy eigenvalues $E_n$ and $|\Omega \rangle$ is the vacuum state.
In the last part of the above expression $E_0$
is the mass of the lightest state contributing to $F(t)$.
The operator $O$ is defined to be
Hermitian, in which case $F(t)\geq 0$ for all $t$. This is also true if we consider the PTP
propagator $F(r)$ instead, where $r$ is the Euclidian distance. 
The matrix defining the spin structure
in the operator $O$ is $\Gamma=i\gamma_5,i\gamma_{i}$ 
for the $\eta_c$ and $J/\psi$ states, respectively, in terms of Hermitian $\gamma_5$ 
and $\gamma_i$. 
At large distances $r$,
the lightest possible modes that couple to the operator $O$
should dominate in $F(r)$. The origin of these can be
light glueballs and, in the dynamical case, the propagation of hadronic 
modes consisting of quarks lighter than the charm quark. 
 Since $F(r)$ is nonnegative for all
$r$, it follows that, when it dominates, $D(r)$ should also be nonnegative in the large distance limit.
The sign of $D(r=0)$, with the above hermiticity condition on $O$, is strictly negative for
the pseudoscalar (and positive for the vector). It follows that in the dynamical case,
where $D(r)$ is dominant at large distances,
$D(r)$ changes sign for the pseudoscalar.
In the quenched case this sign flip occurs if there are glueballs lighter than
the charmonium state studied.
\begin{figure}[t]
\begin{tabular}{cc}
 \epsfxsize=72mm
  \epsfbox{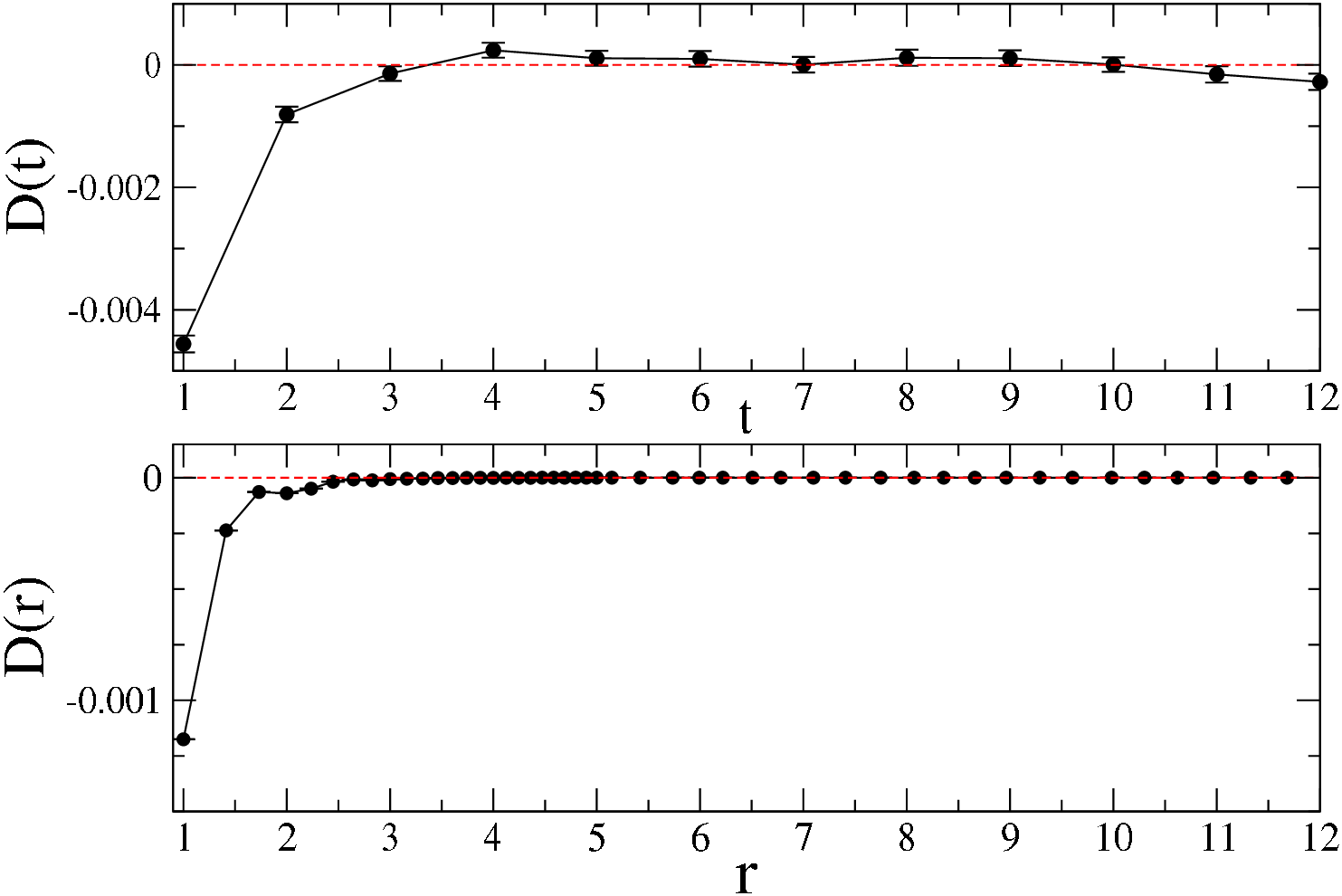}
&
  \epsfxsize=72mm
  \epsfbox{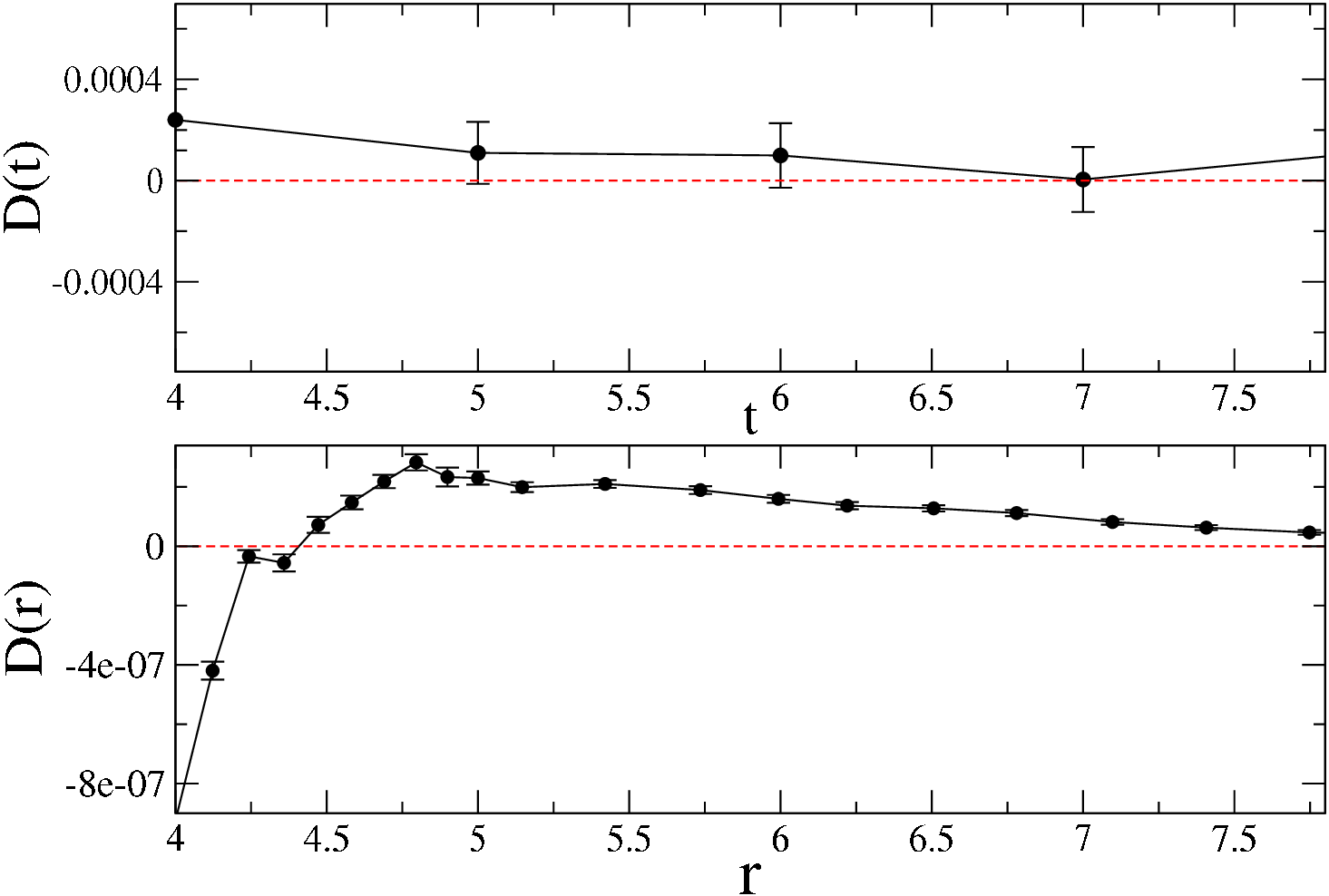}
\end{tabular}
\caption{The comparison of $D(t)$ and $D(r)$ for $\eta_c$ 
for two different ranges of $t$ and $r$ is shown. The results are from the calculation on the dynamical ensemble
at $\kappa_c=0.127$.
The data for $D(r)$ was averaged in small bins in $r$ for $r>5$. }
\label{fig:PP_vs_TS}
\end{figure}
On the lattice, the TTT disconnected propagator is calculated as:
\be
D(t)=c_\Gamma\VEV{L(0)^\star L(t)},\hspace{0.5cm}{\rm where}\hspace{0.5cm}
L(t) = \Tr(\Gamma M^{-1}),
\ee
and $M$ is the charm quark matrix. The trace in $L(t)$ is over the Dirac, color and space indices. For the vector
we have $\Gamma=i\gamma_i$, $c_\Gamma=1$ and for the pseudoscalar $\Gamma=i\gamma_5$,
$c_\Gamma=-1$.  On the other hand, the PTP disconnected propagator is obtained in the following manner:
\be
D(r)=\frac{c_\Gamma}{N_r}\sum_{r=\left|x-y\right|}\VEV{L(x)^\star L(y)},
\ee
where $x$ and $y$ are lattice coordinates,
the sum is over all pairs of lattice points at Euclidean distance $r=\left|x-y\right|$,
$N_r$ is the number of these pairs and there is a trace only over spins and colors
in $L(x)$ (but not over space). 
From the previous studies \cite{fs1,fs2} it is known that the $D(t)$ signal
disappears very quickly around $t=2-3$.
We work with the PTP disconnected propagator instead, since this way we benefit
from both the additional data at noninteger distances 
and the much improved statistics.
The correlator $D(r)$ has from one to three orders of magnitude smaller 
{\it relative} errors than the TTT
disconnected propagator in the region where we have a signal. Figure~\ref{fig:PP_vs_TS}
illustrates this statement by comparing $D(r)$ and $D(t)$ for $\eta_c$
for two different ranges of $r$ and $t$. Both propagators are calculated with $\kappa_c=0.127$ on the dynamical 
lattices from Table~I. 
In the right panel of Fig.~\ref{fig:PP_vs_TS}, the comparison is done on 
a shorter range in order to emphasize the fact that we do have a clear signal
for $D(r)$ in the range where the $D(t)$ signal is completely 
obscured by the noise. 
The result that the $D(r)$ signal is so much better than the one for $D(t)$ can
be explained by the fact that in the calculation of $D(t)$ there are a 
great number of contributions from points, which, although not far from each other
in the $t$ direction, are far in the 4d Euclidean space.
For the disconnected correlator, the noise increases strongly with the distance
and such points contribute nothing to $D(t)$ but noise. This problem is
avoided when working with $D(r)$ instead. We also note that as predicted in
the previous paragraphs, both $D(r)$ and $D(t)$ undergo a sign flip for the $\eta_c$
state.

\section{Fitting the disconnected propagators}

To determine $\delta m$ for the $\eta_c$ we use Eq.~(\ref{dm}),
which means that we have to obtain $\lambda(p^2)$  
from our data for the PTP disconnected propagator. 
In order to fit our data for $D(r)$, we need
a fitting model which satisfies the requirement that the charmonium disconnected propagator
is treated as a composite object,
which has contributions not only from the studied charmonium ground state, 
but also possible effects from
excited charmonium states, states lighter than the charmonioum ground
state, and possibly the $U_A(1)$ anomaly. We also
have to take into account that our data exhibits rotational 
symmetry violations at short distances, due to the finite
lattice spacing. 

To define such a fitting model it is easiest to start from 
the momentum-space description of the disconnected propagator. 
A simplified form which describes its behavior in momentum space is 
\be
D(p^2)= \lambda(p^2)
\left( \frac{\sqrt{A}}{p^2+m_c^2} + \sum_{n=1}^N\frac{\sqrt{A^n}}{p^2+(m^n_c)^2}\right)^2,
\label{eq:fit_full}
\ee
where we have included in the quark loops one ground state, characterized by mass $m_c$ and $N$ excited states with
masses $m_c^n$ (the index $n=1,\dots, N$). 
Here we also make the assumption that the interactions for all states are 
described by the same function
$\lambda(p^2)$.
In the fully quenched case, we model the function $\lambda(p^2)$ as
\be
\lambda(p^2) = U +\frac{f}{p^2+m_g^2}\,,
\label{ql}
\ee
where $U$ stands 
for possible effects of the $U_A(1)$ anomaly, 
so it is negative,
and
$f/(p^2+m_g^2)$ is an effective light glueball term with $m_g$ being the glueball mass. 
We assume that $U$, $f$, $A$, and $A^{n=1\dots N}$
change little for a wide range of momenta and will approximate them with constants 
in our model.
In the 2+1 flavor dynamical case, the expression for $\lambda(p^2)$ could be more complicated. For example,
we need to take into account the existence of both light glueballs and light hadronic modes in order to describe
our data: 
\be
\lambda(p^2) = U +\frac{f}{p^2+m_g^2}+\frac{l}{p^2+m_l^2}+\cdots.
\label{ul}
\ee 
In the above $l$ is a constant and $m_l$ is the  mass of one of the
light hadronic modes. In practice we keep only one light hadronic mode
in the above with an effective mass $m_l$ that (we hope) describes well the long     
distance behavior of the PTP propagator. 

We want to limit the number of free parameters in our model
to as few as possible, since although our data is of much higher quality than in other studies,
it is still difficult to resolve all of the parameters in Eq.~(\ref{eq:fit_full}) 
from the disconnected propagator data. The masses $m_c$ and $m_c^{n=1\dots N}$, for example,
can be determined from fits to 
the TTT connected charmonium propagator $C(t)$,
and then be used as constants in our fits. We obtain the ground state mass $m_c$ with a very small fitting error, but
the excited states masses $m_c^{n=1\dots N}$ are less well known.  We can also determine with varying degrees of precision
the constants $A$ and $A^{n=1\dots N}$
from fits to $C(t)$, since they are proportional to the amplitudes of the
ground and the excited states, respectively. 
We relate the PTP amplitude $A$ to the corresponding TTT amplitude as follows:
We Fourier transform the ground $C(p^2)$ propagator and then
integrate over space:
\be
\int d^3x\int\frac{d^4p}{(2\pi)^4}\frac{A}{p^2+m_c^2}e^{ip r}
= \frac{A}{2m_c}e^{-m_ct}.
\ee
The right-hand side in the above is the TTT propagator, which implies a relation between the amplitude of the ground state in $C(t)=A_te^{-m_ct}+\cdots$, denoted by $A_t$,
and  the factor $A$:
\be
A= 2m_cA_t\,.
\label{a}
\ee

Another parameter that we can fix in our model using prior knowledge is the glueball mass
$m_g$. We use the results for the lightest $0^{-+}$ glueball from Ref.~\cite{glue}, namely we set $m_g =2563$ MeV,
which is the value extrapolated to the continuum limit.
We use this value for all of our lattice spacings, since in Ref.~\cite{glue} it was found that the glueball mass
does not vary much at fine lattice spacings and is compatible with the continuum extrapolated result.   
On the dynamical ensembles we have to take into account also the contribution
of the light hadronic modes, and preferably we want also to set the mass $m_l$ to 
an appropriate constant. In our previous work \cite{mine,mine1} we found that the long distance behavior
of the PTP pseudoscalar propagator on the dynamical ensemble 
can be fitted well with a light state of mass $ am_l\sim0.42$. This is very close 
to 
the physical mass of the $\eta^\prime$ of 958 MeV; thus we fix $m_l$ to the mass of the $\eta^\prime$. 
Although there are states lighter than the $\eta^\prime$ 
contributing as well (such as the $\eta$, multipion states {\it etc.}),
this approximation is probably satisfactory, because
the modes lighter than the $\eta^\prime$ would mix even less with the heavy $\eta_c$ state,
and we cannot 
distinguish their signal at our level of statistics. Thus the only free parameters in our model remain
$U$, $f$, and $l$ (the last one is present only in the dynamical case).

The summary of our fitting strategy in the simpler fully quenched case is
as follows:
\begin{itemize}
\item On a given lattice ensemble we calculate the TTT connected propagator of 
the $\eta_c$ state. From fits to it with the asymptotic form 
$C(t)=A_te^{-m_ct}+ \sum_{n=1}^NA_t^n e^{-m_c^n t}$ we determine 
$m_c$, $m_c^{n=1\dots N}$, $A_t$, and $A_t^{n=1\dots N}$. Using Eq.~(\ref{a}) with
the substitution $m^{(n)}_c\rightarrow \sqrt{2(\cosh (m^{(n)}_c) -1)}$ in order
to take into account lattice discretization effects, we obtain $A$ and $A^{n=1\dots N}$.
We use the central values of all of the above parameters as constants in our model
function  Eq.~(\ref{eq:fit_full}).
\item We also fix the parameter $m_g$ in Eq.~(\ref{ql}), using prior knowledge.
\item In Eq.~(\ref{eq:fit_full}) we replace $p^2$ with $\sum_i 2(1-\cos(p_i))$  and
all the masses with the appropriate expression as done in the first item
above for $m^{(n)}_c$, to account for 
the lattice discretization effects.
Equation~(\ref{eq:fit_full}) can be rewritten in a form
linear in the two parameters $U$ and $f$, which is more convenient for fitting
purposes:
\be
D(p^2) = UT_1(p^2) + fT_2(p^2).
\ee
Next, we do a discrete Fourier transformation of $T_{1,2}(p^2)$ on a lattice of an appropriate size
and obtain the functions $T_{1,2}(r)$ at discrete values of $r$.
We tabulate $T_{1,2}(r)$ at each distance $r\leq15$. This range of $r$ is sufficient, since our 
signal is too noisy for $r>15$.
Thus, using the linear model 
\be
D_{\rm fit}(r) = UT_1(r) + fT_2(r),
\ee
we fit our data for the PTP disconnected charmonium propagator $D(r)$ in order
to extract $U$ and $f$.
\item With the fit values of $U$ and $f$ at hand, we determine $\lambda(-m_c^2)$ and
 $\delta m$ from Eqs.~(\ref{dm})~and~(\ref{ql}).
\end{itemize}
The fitting procedure in the 2+1 flavor dynamical case is quite similar. In addition there
we  
fix appropriately the light mass $m_l$ and use a fitting form with three tabulated terms:
\be
D_{\rm fit}(r) = UT_1(r) + fT_2(r)+ lT_3(r). 
\ee
After we extract the parameters $U$, $f$, and $l$, we solve again for $\delta m$ with
the appropriate $\lambda(-m_c^2)$ from Eq.~(\ref{ul}).

The error due to the assumption that the participating masses and the amplitudes $A$
and $A^{n=1\dots N}$ in our model are constants is discussed in Sec.~VI.
The success of our fitting model depends on 
how well it approximates the interaction dynamics on the lattice and on the quality of our
data. An essential part of the construction of the fitting model for the PTP disconnected propagator 
turned out to be 
the number of excited
states $N$ which we have to include. The excited states have larger contributions to the
disconnected propagator at a given distance than they have in the case of the connected 
propagator. This means that a good knowledge of the spectrum of excited states is required
in order to fit the disconnected propagator.
In the disconnected diagram with net propagation distance $r$ the connected $\eta_c$ propagates on average less than
half the distance before annihilating and again less than half after reappearing.
This means that a reliable fit to the connected propagator yields masses and amplitudes which
we can use in fits to the disconnected propagator at least at twice the distance, where the signal may be 
too noisy. Thus, we are limited to fits of the connected propagator with $t_{\rm min}=2$ or 3
and use the extracted parameters in the disconnected fits from $r_{\rm min}=4$ or 6.     

To obtain the charmonium spectrum from the fits to the connected propagator,
we employ a fitting model which forces the 
splittings between the states to be positive, essentially creating a "tower of states". 
The priors for the logarithms of all the splittings are the same ({\it i.e.}, we assume 
the states are equidistant as a first approximation) and so are their widths. 
We used a set of different values for the splitting priors 
and their widths to check the stability of the resulting spectrum. We 
found that the extracted masses were stable for a relatively wide range
of priors. Our best fits have the splitting priors in the range of 570--770 MeV.
Still, this approach is not intended to provide a reliable determination
of the masses of excited quarkonium.  Rather, we use this heuristic model
to understand the role played by excited states in our result.
  
In our fitting model  
we also require (through the use of priors) that the amplitudes of the excited states are no larger than the 
ground state amplitude. Without this restriction, we noticed that the amplitudes
of the excited states grow noticeably larger than that of the ground state
(where we define this to mean difference larger than $1.5 \sigma$). The assumption that this should not happen
stems from general considerations: The wave function at the origin
is smaller for an excited state than the ground state, simply because the excited state wave function spreads out more.  This is a characteristic of
nonrelativistic potential models. There is also some support from
experiment: The decay constant of $\psi^\prime$
is smaller than that of the $J/\psi$ [PDG values: 279(8) MeV {\it vs} 411(7) MeV, respectively].
Lattice studies of 
the light-light \cite{l-l}  and heavy-heavy \cite{h-h}
meson sectors also confirm this expectation. 
But other lattice calculations, such as in Ref.~\cite{h-l} for the heavy-light meson case,  
show the first excited state decay constant growing larger than the ground state one. 
A possible explanation for this discrepancy in the last study is that
the contributions of the neglected higher excited states became
lumped into the amplitude of the first excited state.

In our fitting model, the requirement that the amplitudes of the excited states do not grow much  
larger than the ground state amplitude,
we hope, prevents the ``clumping'' of states with similar masses into an
effective state with a large effective amplitude. 
However, since we use only a Gaussian prior to constrain the logarithm of the excited
state amplitudes, it still often happens, depending on the number of
states included, that the highest state in the fit and sometimes the
next highest end up with large amplitudes.  We interpret this outcome
as a clumping of multiple unresolved states of similar mass.  How we
count them affects our result for $\delta m$.
To compensate for this effect we represent the resulting contribution
as a sum of states of similar mass:
\be
 \frac{A^N}{(p^2 + (m_c^N)^2)} \approx \sum_{k=1}^M \frac{A^\prime}{(p^2 + (m_c^N)^2)}.
\ee 
where $M\approx A^N/A^\prime\equiv A^N_t/A_t$ (the ratio of the highest excited state amplitude to the ground state one).
Each of the states in the sum above should contribute $\sqrt{A^N/M}/(p^2 + (m_c^N)^2)$ in Eq.~(\ref{eq:fit_full}).
Since there are $M$ of them, the contribution of the effective $N$-th state in Eq.~(\ref{eq:fit_full}) is modified
to $\sqrt{A^NM}/(p^2 + (m_c^N)^2)$. In effect, we multiply the amplitude $A^N$ by the number $M$ 
of its contributing states before using it in Eq.~(\ref{eq:fit_full}) in order to correctly account 
for the possibility that the $N$-th state is an effective one.  
Note that increasing the multiplicity in this way always decreases $\delta m$.
Thus in our analysis of systematic effects, we explore the sensitivity of our
result to this assumption.

\section{Tuning the charm quark mass}
\begin{figure}[t]
  \includegraphics*[width=8cm]{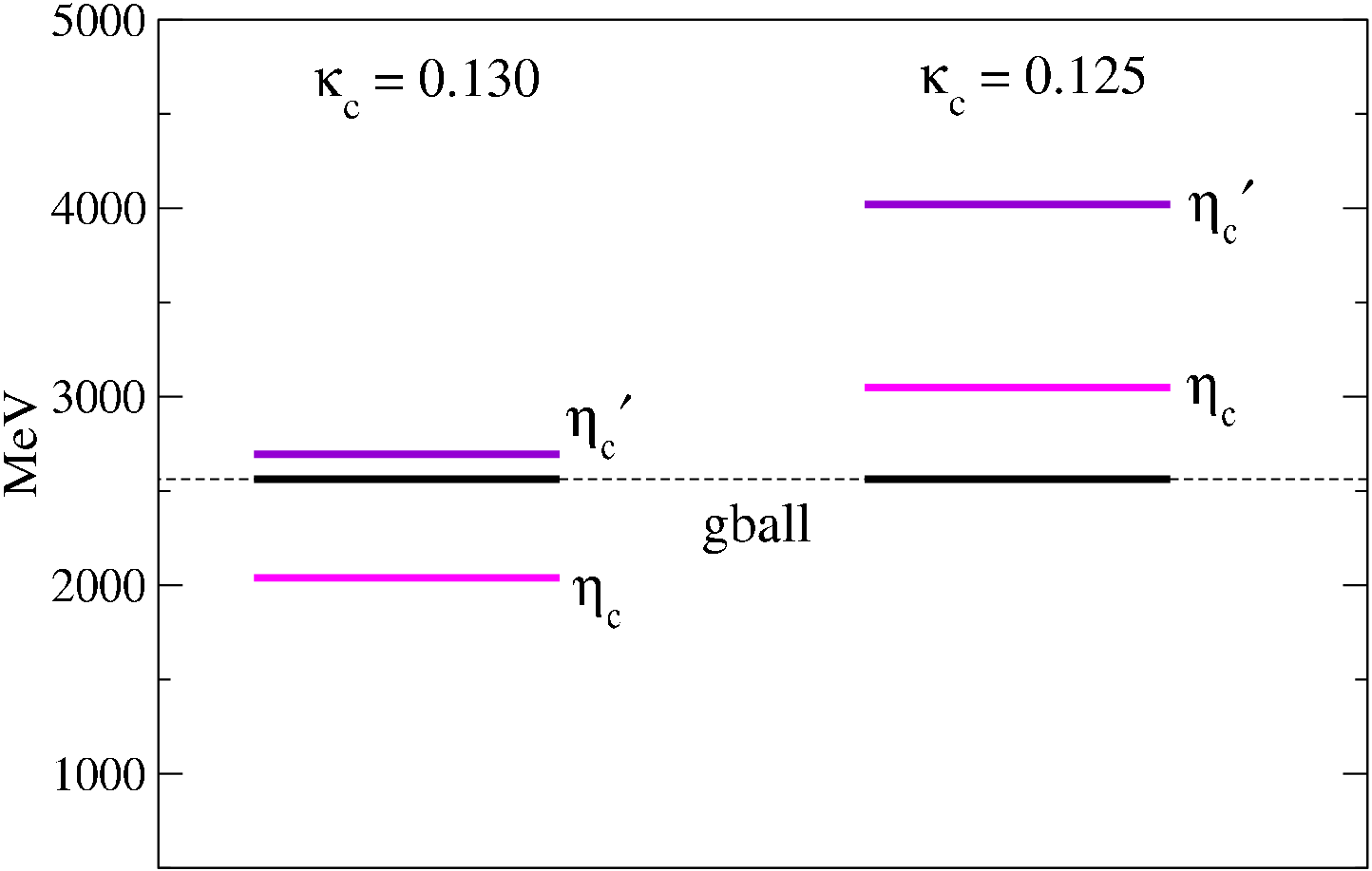}
  \includegraphics*[width=8cm]{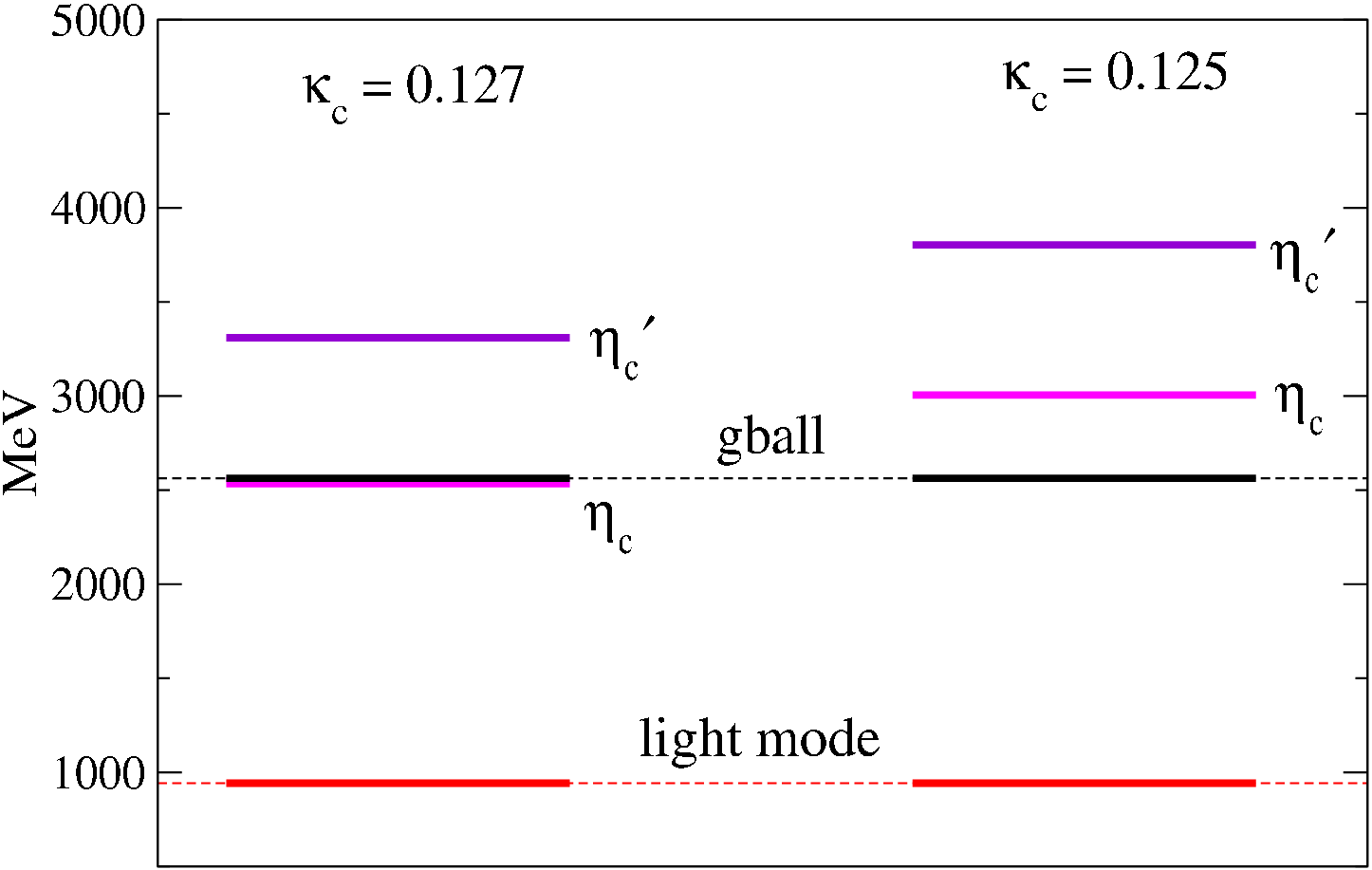}
\caption{(Left panel) Relative positions of the charmonium states with respect to the
$0^{-+}$ glueball mass with $m_g=2563$ MeV on the quenched superfine ensemble,
for different $k_c$, are shown.
(Right panel) Same for the dynamical ensemble.
}
\label{fig:etacposition}
\end{figure}
As we already stated in the Introduction, we use clover fermions to generate the
connected and disconnected charmonium propagators. This means that we have 
to tune the hopping parameter, $\kappa_c$,
to correspond to the physical charm quark mass. In our preliminary work \cite{mine,mine1} we 
tuned $\kappa_c$ using the kinetic mass of $D_s$ {\it i.e.}, we used the Fermilab 
interpretation of the clover fermions. For the fully quenched ensembles this tuning was
rather approximate. In this work we adopt a different approach,
namely, we tune $\kappa_c$ by matching the rest mass of the $\eta_c$ to its physical mass instead
with an accuracy of several percent. 
Our current approach is more appropriate for the purposes of determining $\delta m$,
since it positions the charmonium $\eta_c$ state correctly  with respect
to the lightest glueball with which it might mix. This is important, because, depending
on whether the mass of the charmonium state is heavier or lighter than the lightest glueball,
$\delta m$ might change in absolute value or even undergo a sign flip. 
Figure~\ref{fig:etacposition}
illustrates this statement by showing the masses of the $\eta_c$,  $\eta_c^\prime$ 
and the lightest $0^{-+}$ glueball on the superfine quenched (left) and 
the dynamical fine ensembles (right),
for values of $\kappa_c$ obtained by our previous and 
current tuning methods. In the quenched superfine case, for example, using the value $\kappa_c=0.130$
from the kinetic mass tuning of $D_s$, gives a rest mass of the $\eta_c$ lighter than its physical value and
lighter than the lightest glueball.
This implies that their mixing will "push" the $\eta_c$ mass to lower values
on the lattice. However, if the $\eta_c$ rest mass assumes its correct physical value
(achieved at $\kappa_c=0.125$),
the effect of the glueball mixing would be exactly the opposite.  
We conclude that 
although the kinetic mass $\kappa_c$ tuning
is the correct method in cases when we want to determine various mass 
spectrum splittings
in the charmonium system,  for our study the appropriate method  
is to tune the charm quark mass using the rest mass of the charmonium state we
are interested in. (An alternative method which could render both the charmonium splittings and the
rest masses correct, is to have different values for the spatial and temporal hopping parameters,
 a strategy which we do not employ here.) 

\section{Results and conclusions}
\begin{table}[t]
\begin{footnotesize}
\begin{tabular}{l}
\begin{tabular}{lcccccccc}
\hline\hline
\#states & $m_{\eta_c}$ & $m^1_c$ & $m^2_c$ & $m^3_c$ & $m^4_c$ & $m^5_c$& $m^6_c$&$\chi^2/$DOF \\
\hline
4& 1.3708(2)& 1.67(3) & 1.98(6) & 2.59(5)& $\cdots$&$\cdots$&$\cdots$&1.6\\
5& 1.3708(2)& 1.68(3) & 1.98(7)& 2.56(5)&2.81(13)&$\cdots$&$\cdots$&1.6\\
6& 1.3708(2)& 1.67(3)& 1.94(6) & 2.16(9)& 2.42(10)&2.67(11)&$\cdots$&1.4\\
7& 1.3709(2)& 1.67(3)&1.93(7)&   2.13(8)& 2.34(9)&2.54(9)&2.74(9)&1.3\\
\hline\hline
\end{tabular}\\
\\
\begin{tabular}{lccccccc}
\hline\hline
\#states &$A_t$ & $A^1_t$& $A^2_t$& $A^3_t$& $A^4_t$& $A^5_t$& $A^6_t$\\
\hline
4& 0.589(3) & 0.58(19) & 1.63(34) & 5.64(31)&$\cdots$&$\cdots$&$\cdots$\\
5& 0.589(3)&0.59(19)&1.58(35) & 4.81(45)& 0.91(36)&$\cdots$&$\cdots$\\
6& 0.589(3)& 0.56(18)&1.06(33)& 0.93(35)&0.86(34)& 4.55(50)&$\cdots$\\
7& 0.588(3)&0.56(17)&0.90(31)& 0.95(35)&0.95(37)&0.97(38)&3.74(60)\\
\hline\hline
\end{tabular}\\
\\
\begin{tabular}{lll}
\hline\hline
\#states & $\delta m$ [MeV] &  $\delta m^{\rm corr}$ [MeV]\\
4& -3.61(24)& -1.59(15)\\
5& -3.31(29) &  -1.62(15)\\
6& -2.74(24) & -1.75(20)\\
7& -2.45(22)& -1.88(18)\\
\hline\hline
\end{tabular}\\
\end{tabular}
\end{footnotesize}
\caption{ Masses (top) and amplitudes (middle) in lattice units 
extracted from fits to 
the connected propagator on the QF ensemble with different numbers of states
(4, 5, 6, and 7) are shown. The lowest part of the table shows the mass shift $\delta m$, 
calculated using the results from the upper two parts of the table. Also shown is the "corrected" mass shift
$\delta m^{\rm corr}$, which is obtained using the systematic error estimation method described at the end of Sec.~IV. }
\label{tab:QFres}
\end{table}
\begin{table}[t]
\begin{footnotesize}
\begin{tabular}{l}
\begin{tabular}{lccccccccc}
\hline\hline
Ensemble & $m_{\eta_c}$ & $m^1_c$ & $m^2_c$ & $m^3_c$ & $m^4_c$ & $m^5_c$  & $m^6_c$ & $m^7_c$ & $m^8_c$\\
\hline
QF &   1.3708(2)  &1.67(3)&1.94(7)& 2.16(9)  &2.42(10)  &2.67(11)&$\cdots$ &$\cdots$&$\cdots$\\
QSF & 0.9734(2) &1.22(2)&1.41(7)&1.63(9)&1.87(10)& 2.10(10)&$\cdots$&$\cdots$&$\cdots$  \\
DF&1.2749(4) & 1.58(3)& 1.88(7)&2.08(9)&2.26(8)&2.43(8)&2.62(10)&2.83(11)&3.05(12) \\
\hline\hline
\end{tabular}\\
\\
\begin{tabular}{lccccccccc}
\hline\hline
Ensemble &$A_t$ & $A^1_t$& $A^2_t$& $A^3_t$& $A^4_t$& $A^5_t$& $A^6_t$& $A^7_t$& $A^8_t$\\
\hline
QF & 0.589(3) &0.56(18)&1.06(33)& 0.93(35)&0.86(34)& 4.55(50)&$\cdots$&$\cdots$&$\cdots$\\
QSF & 0.270(2)&0.32(10)&0.39(13)&0.33(13)& 0.33(13)&4.44(24)&$\cdots$&$\cdots$&$\cdots$\\
DF& 0.822(7)& 0.84(21)&1.00(37)&1.30(47)&1.40(49)&1.30(47)&1.17(43)
&1.07(40)&0.98(38)\\
\hline\hline
\end{tabular}\\
\end{tabular}
\end{footnotesize}
\caption{Parameters extracted from fits to the TTT connected $\eta_c$ propagator $C(t)$ for
each ensemble are shown. Masses and amplitudes are in lattice units. The $\chi^2/$DOF for the fits 
is 1.4, 2.2, and 1.4 for the QF, QSF, and DF ensembles, respectively. The $t_{\rm min}$
for these fits is 2, 3, and 2. The central
values of the masses and amplitudes are used as constants in the fitting model for the disconnected $\eta_c$ propagator.
}
\label{tab:param}
\end{table}
In this section we present our results for the mass shift $\delta m$ due to the
contribution of the disconnected diagrams for the $\eta_c$ on all of the ensembles 
from Table~\ref{tab:lat}.
We use 72 $Z_2$ random sources per lattice with spin and color dilution  
to compute the disconnected propagators
on all of our ensembles (which means there are $72\times 12$ quark matrix 
inversions per lattice performed). 
To explore the systematic effects which arise in the determination of 
$\delta m$ due to our incomplete knowledge of the charmonium spectrum, we studied
in more detail the data from the quenched fine ensemble. We fitted the connected propagator
using 4, 5, 6, and 7 states and used the extracted masses and amplitudes to fit
the disconnected propagator $D_{\eta_c}(r)$, as described in Sec.~IV, in each case.

Table~\ref{tab:QFres} shows the results from the fits to the connected propagator 
in its upper and middle subtables. All of the fits were performed on the same time range
($t=2-45$). We see that the $\chi^2$/DOF improves with adding more states to the fitting model
and the amplitudes of the excited states become smaller at the same time. The extracted masses
of the ground
and the two lowest excited states appear to be quite independent of the number of states included in the fit.
The third and higher excited states on the other hand 
do depend on the number of states. 
This is not very surprising; these states
are much more difficult to extract and are likely to be effective states. Their amplitudes also are more likely to
grow large, lending support to this interpretation.  

The effect of the number of states in each
fit on $\delta m$ is shown in the lower part of Table~\ref{tab:QFres}. We also show there the
corrected value of the mass shift $\delta m^{\rm corr}$, which is obtained by modifying 
the amplitudes that are significantly larger than the ground state one 
({\it i.e.}, the amplitudes of the second and third excited states
for the 4- and 5-state
fits and the highest excited state one in the case of 6- and 7-state fits), in the manner 
described at the end of Sec.~IV. The difference
between $\delta m$ and $\delta m^{\rm corr}$ gives some idea of the systematic error. 
This systematic error grows when there are fewer states in the fitting model, possibly because
the charmonium spectrum is not well represented by it over the fitting range. This effect is
also signaled by a growing   
$\chi^2$/DOF in these cases. 
It is  encouraging that $\delta m^{\rm corr}$ is very
consistent between the different fits. 

Among these results for $\delta m$, the one we
consider best is the one obtained using the 6-state fit to the connected propagator, for two reasons. 
First, this is the fit with the fewest
excited states to achieve $\chi^2/$DOF
below 1.5, a value we consider to be on
the boundary between "good" and "bad" fits. Second, the excited state spectrum is close to the picture
where all amplitudes but that of the highest included state are not much larger than the ground state amplitude. 
We expect the highest excited state to reflect the fact that we work with finite number of states
in the fitting model,
and thus likely to have a large effective amplitude. 

We apply the same criteria when we repeat the whole calculation on the
rest of the ensembles in this study. We adopt the results for $\delta m$ from the best of our fits as the final 
answer and report the difference between  $\delta m$ and $\delta m^{\rm corr}$ as an asymmetric systematic error.  
Results for all three ensembles are summarized in
Table~\ref{tab:param}.  The number of states in the preferred fit is
chosen according to the criteria described above, resulting in six
states for the quenched ensembles and nine for the dynamical one.  For
the quenched superfine ensemble however we did not obtain a good
$\chi^2/$DOF for $t_{\rm min}=2$ or 3, even when we included more than
six states.

\begin{figure}[t]
  \includegraphics*[width=8cm]{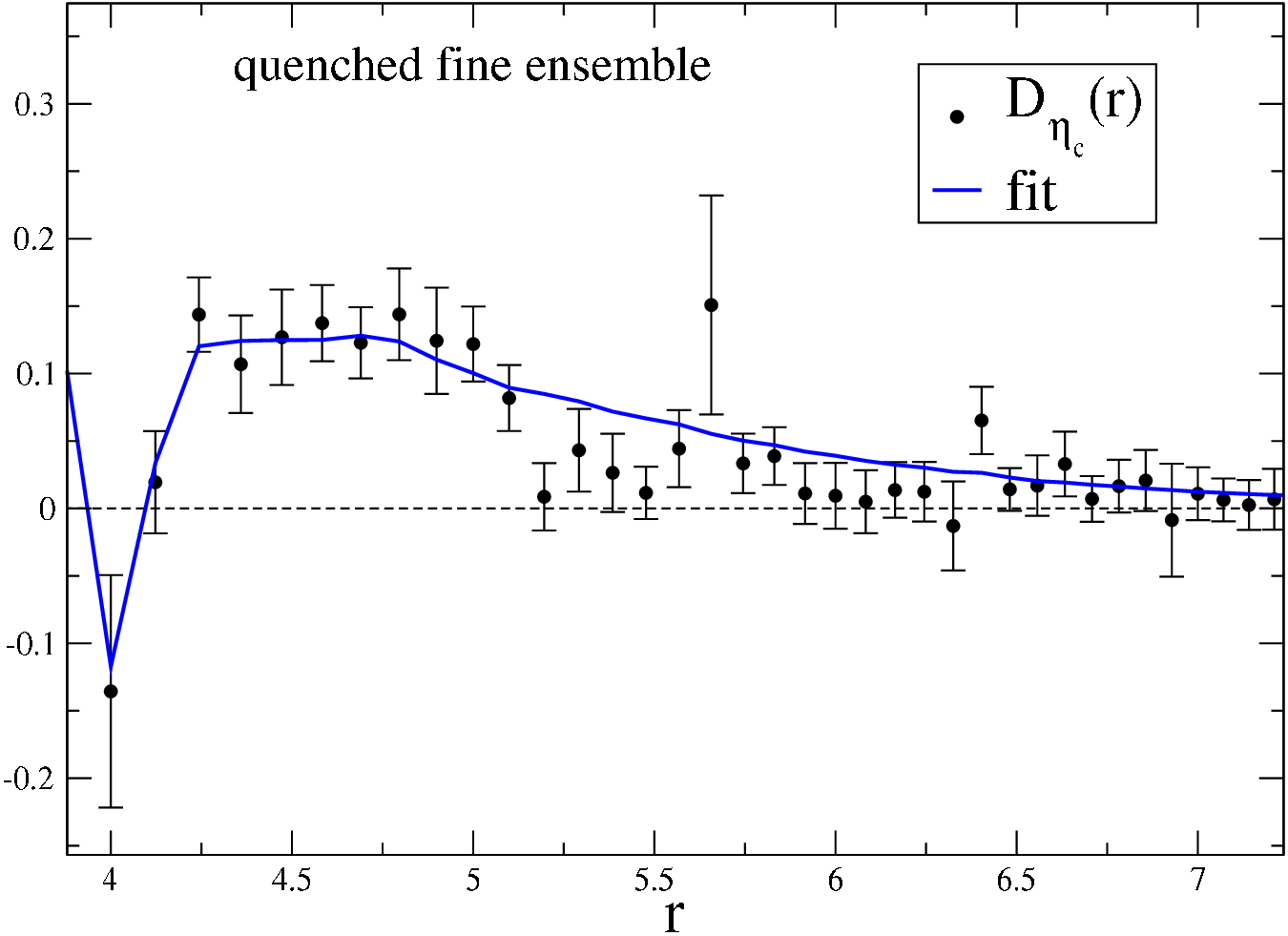}
  \includegraphics*[width=8cm]{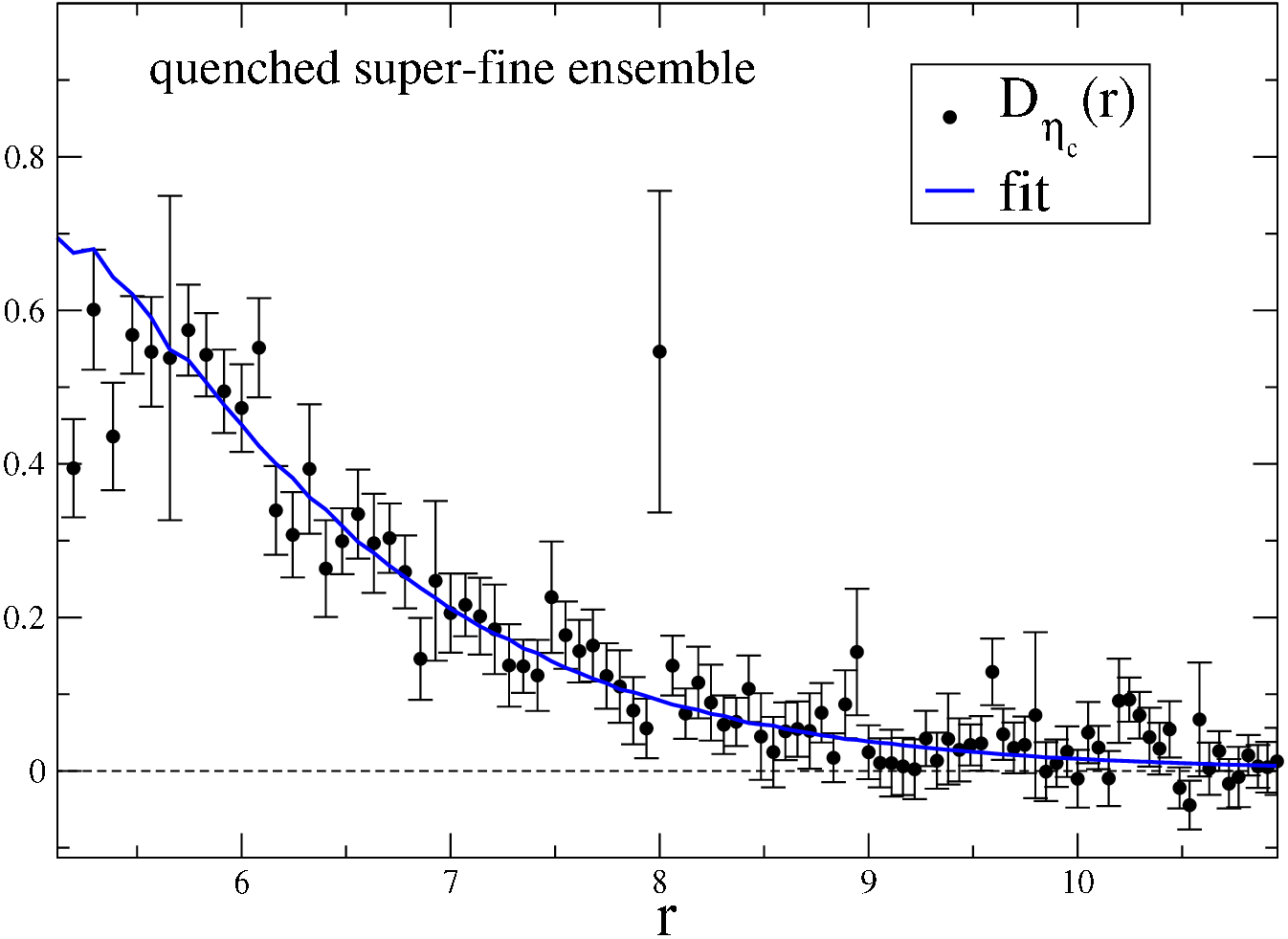}
\caption{(Left panel) The $\eta_c$ disconnected propagator {\it vs}
the Euclidean distance $r$ in lattice units for the quenched fine ensemble at $\kappa_c=0.120$
is shown.
Fluctuations in the data larger than the statistical errors are due to rotational symmetry 
violations.%
\protect\footnote{A Fourier transform of a 
     monotonic function performed on a discrete lattice (which violates 
     rotational symmetry) may exhibit
     nonmonotonic "fluctuations".}
(Right panel) Same for the superfine quenched ensemble at $\kappa_c=0.125$.}
\label{fig:qu_fits}
\end{figure}

\begin{figure}[t]
  \includegraphics*[width=8cm]{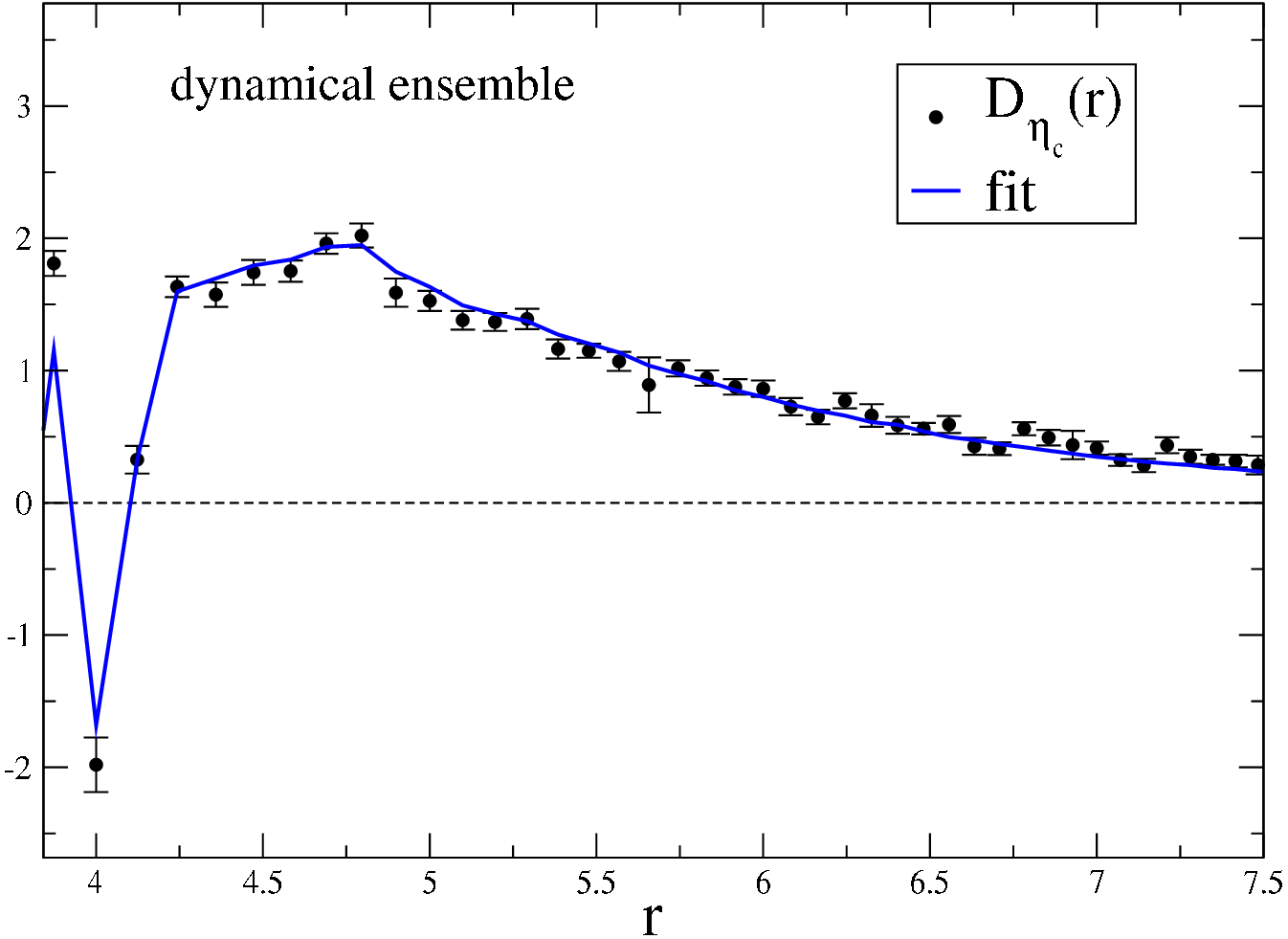}
  \includegraphics*[width=8cm]{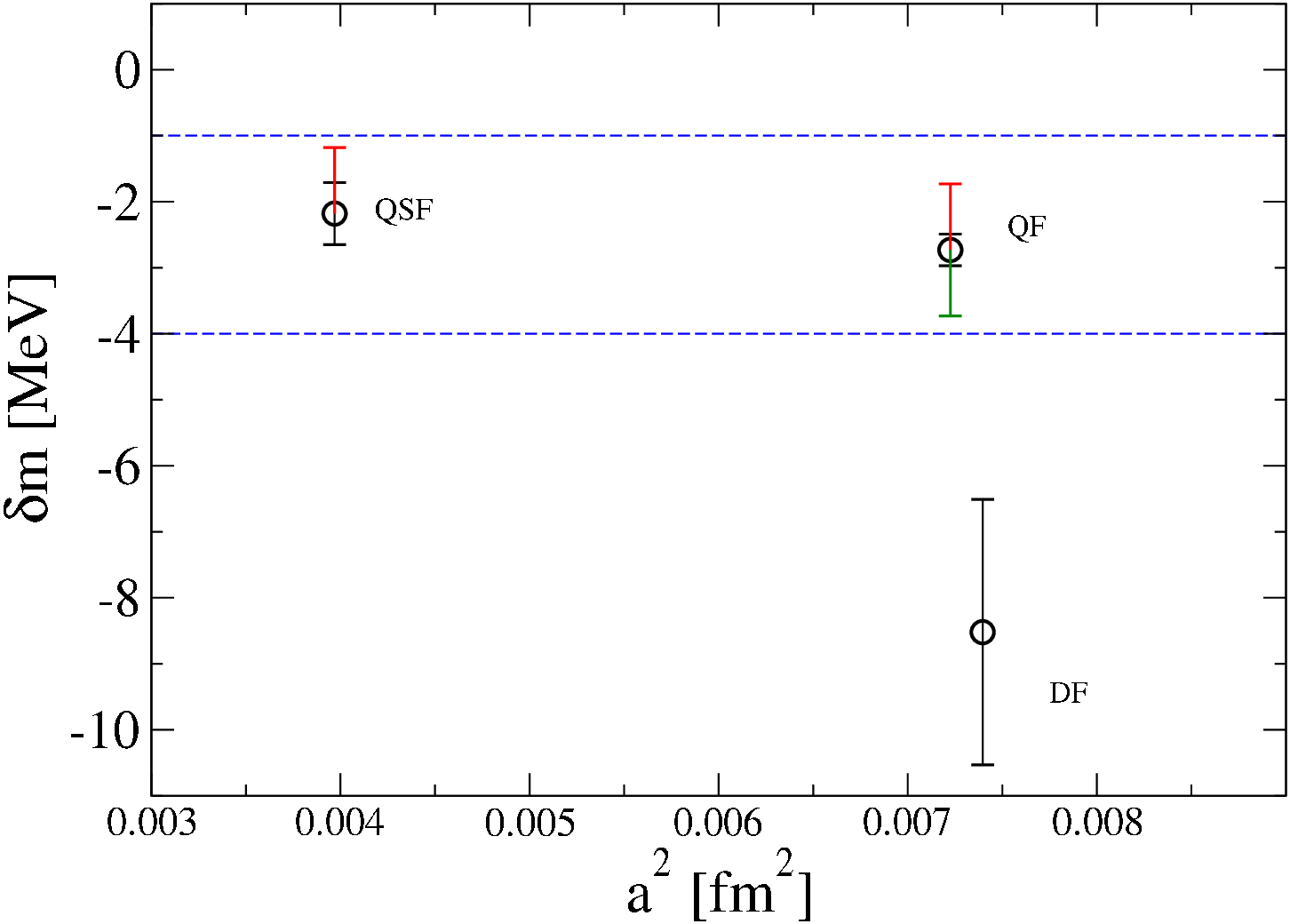}
\caption{(Left panel) The $\eta_c$ disconnected propagator {\it vs}
the Euclidean distance $r$ in lattice units for the dynamical fine ensemble at $\kappa_c=0.125$ is shown.
(Right panel) The mass shift $\delta m$ for the QF, QSF, and
DF ensemble. The black error bars denote the statistical errors for the QF and QSF 
results. For the DF ensemble, the black error bar also includes the error due to specific
assumptions in the fitting model as described in the text. The asymmetric
red error bar originates from the systematic error due to our incomplete knowledge
of the excited charmonium spectrum. The asymmetric green error bar is due to the $\kappa_c$ mistuning
in the QF case. The band between the two slashed blue lines encompasses the most likely range of
$\delta m$ based on the quenched data.
}
\label{fig:dyn_fit}
\end{figure}

To complete our results
 for each ensemble, we fit our $D_{\eta_c}(r)$ data, as described in Sec.~IV,
with the values for the model parameters taken from Table~\ref{tab:param}, respectively. 
The glueball mass in our model is the constant $m_g=2563$ MeV, and in
the dynamical case we also fix $m_l=958$ MeV. Figures~\ref{fig:qu_fits} (both panels) and \ref{fig:dyn_fit} (left panel)
show our results for the $\eta_c$ disconnected propagator for each of the studied ensembles
and the best fits to the data. In Table~\ref{tab:results} we give the fitting parameters, ranges,
$\chi^2$ per degree of freedom  
and our final result for the mass shift $\delta m$ for each ensemble. 

All of our results
consistently show $\delta m<0$ (meaning an increase on the $\eta_c$ mass).
It is notable that this is opposite the prediction 
of perturbation theory
\cite{hisq}.  
Even without a quantative calculation of the mass shift $\delta m$ one might
expect that it will be negative. 
First,  according to the mixing models \cite{Feldmann:1998vh}, 
if the $U_A(1)$ anomaly has an effect on the $\eta_c$, it should make  
its mass larger. In our model this is reflected
in the fact that we obtain $U<0$. Second, the light glueball  happens to be lighter than
the $\eta_c$, and when they mix, similarly, the mass of the $\eta_c$ 
is pushed up. 
This is reflected in our finding that $f/(-m_c^2+m_g^2)<0$.
The same is valid for the effect of light hadronic modes on $m_f$ in the dynamical case.
The mass shift itself is
similar for the two quenched ensembles: $\delta m=-2.74(24)$ and $-2.18(47)$ MeV for the fine 
and the superfine one, respectively, where the errors are statistical only. We estimate the systematic effects stemming from 
our limited knowledge of the charmonium excited states as described in Sec.~IV, to be around 
1 MeV, applied in the direction of decreasing the absolute value. This estimate is based on the
difference between the above values for $\delta m$ and their corresponding $\delta m^{\rm corr}$, the latter being $-1.75(20)$ MeV and $-1.15(28)$ MeV, 
respectively. The quenching of the light quarks 
is, of course, another source of systematic error which might not be
negligible. 

The dynamical ensemble 
yields a larger value for the $\eta_c$ mass shift:
$\delta m =-8.52(24)$~MeV (the error is statistical only). 
We did not determine the systematic effects of the excited states
(through $\delta m^{\rm corr}$) in this case,
because the amplitudes of the excited states in the connected propagator 
did not grow much larger than the ground state one ({\it i.e}, they differ less than $1.5\sigma$).
This is not too surprising considering 
that we already use 8 excited states in our fit to the connected propagator in order to
achieve a good $\chi^2/$DOF.
The discrepancy in the determined values of $\delta m$ in the quenched and the dynamical cases
is most likely due to much larger
systematic errors (other than the excited states contribution) in the latter case. 
For example, our simplified model 
does not account for the complications due to the  mixing with the open charm threshold.
Above this threshold there are numerous close-lying open-charm states.  
These discretized (in finite volume) continuum states do not fit the tower-of-states
model: At our lattice size their level spacing is approximately ten
times smaller than the typical level spacing in our tower-of-states
model, and they have a degeneracy that grows rapidly with S-wave phase
space.  Further, their amplitudes are likely to be much smaller than
the amplitudes of the ``bound'' states.  They would appear as clumped,
effective states in the tower-of-states model. To the extent they are
important in our analysis, correcting for their clumping would tend to
reduce the absolute value of $\delta m$. 

In Table~\ref{tab:results}
we also present separately the contribution of the anomaly $\delta m_U$ to the total mass shift for each of the ensembles.
The values of $\delta m_U$ are roughly the same as the one predicted 
in Ref.~\cite{Feldmann:1998vh} using mixing models.
Our numerical results show that the contributions in MeV of the $U_A(1)$ anomaly 
is about half to $2/3$ of the final value. The effect of mixing of the $\eta_c$ 
with light hadronic modes in
the dynamical case is much smaller than $1$~MeV and is practically negligible. This is not 
unexpected considering the large mass
difference between them.
Figure~\ref{fig:dyn_fit} (right panel) shows the values of $\delta m$ for each ensemble.

In addition to the statistical error in these values, another source of uncertainty 
in $\delta m$ is the $\kappa_c$ tuning.
We have negligible $\kappa_c$-tuning errors for the superfine quenched (and the dynamical) case where the tuning is
done to $1\%-2\%$, but 
the $\eta_c$ mass in the case of the quenched fine ensemble is about 7\% heavier than the physical one. 
This leads to an asymmetric correction to the $\delta m$ in the case of the quenched fine ensemble 
by about 1 MeV in the direction
of increasing its absolute value. We obtained this correction by assuming that the parameters $U$ and $f$ change negligibly 
for small mass fluctuations and there is only an explicit dependence on $m_c$ in Eq.~(\ref{dm}). 
Then we equate the systematic error with  the
difference in $\delta m$ when we use the physical and the measured value of the mass of the $\eta_c$. 

Finally, the last source of uncertainty in $\delta m$  
originates from the assumption that the masses and amplitudes in the fitting model Eq.~(\ref{eq:fit_full})
are constant, when in reality we know them up to their statistical errors. We estimate the effect of this assumption
by varying these masses and the amplitudes within their statistical errors when fitting the disconnected propagator
and recalculating $\delta m$. 
We found that the resulting error for the quenched ensembles is within
the statistical uncertainty and we neglect it in the final error budget for these ensembles. For the dynamical ensemble this error turned out to be 2 MeV. In this case we add it to the statistical error for this ensemble in Fig.~\ref{fig:dyn_fit} (right panel).  

We mentioned in the Introduction that we equate the effects of the disconnected diagrams
on the hyperfine splitting in charmonium with the mass shift they induce in the 
$\eta_c$ only. In other words, we ignore the possible mass shift they cause in the $J/\psi$.
We base this approximation on our attempt to estimate the effect on the vector using a
fitting procedure similar to the one we used for the $\eta_c$.
Our data for the vector is more noisy and the signal in the PTP propagator dies out 
at shorter distances than in the pseudoscalar case, due to the larger mass of the $J/\psi$. 
We found the effects of the disconnected diagrams for the vector is much smaller than 1 MeV 
 and thus, they are within the statistical error of $\delta m$ for $\eta_c$.
 
In conclusion, based on our results for the mass shift in $\eta_c$ in the quenched case, 
the charmonium hyperfine splitting is decreased by 1 -- 4 MeV when we take into account the disconnected 
diagrams. This range is represented visually by the band between the slashed blue lines 
in Fig.~\ref{fig:dyn_fit} (right panel). In this final range for $\delta m$ we ignore the dynamical
result on the basis of its much larger and  much less reliably estimated systematic effects, 
which require further study.  
\begin{center}
ACKNOWLEDGMENTS\\
\end{center}
We are grateful to Andreas Kronfeld and Peter Lepage for helpful comments.
Computations for this work were carried out in part on
facilities of the USQCD Collaboration, which are funded by
the Office of Science of the United States Department of Energy, and in part at CHPC (Utah).
LL and CD are suppoterd by the National Science Foundation and the United States Department of Energy.
\begin{table}[t]
\begin{tabular}{lccclccc}
\hline\hline
Ensemble & $U\times 10^{3}$ & $f\times 10^3$ & $l\times10^4$ & $\delta m$ [MeV]& $\delta m_U$ [MeV]& $r_{\rm min}-r_{\rm max}$ & $\chi^2/{\rm DOF}$ \\
\hline
QF & $-2.31(36)$& 1.00(14)& $\cdots$ & $-2.74(24)^{+1.0}_{-1.0}$ & $-1.81(28)$&$4-7$ & $36/32$\\
QSF &$-6.53(35)$& 0.240(65) & $\cdots$ & $-2.18(47)^{+1.0}_{-0}$ &$-1.01(53)$  &$6 -10$& $60/63$ \\
DF & $-5.29(41)$& 2.28(15)& $0.413(52)$ & $-8.52(24)^{+2.0}_{-2.0}$&$-4.45(21)$ & $4-10$ &$111/82$\\
\hline\hline
\end{tabular}
\caption{Fitting results for the disconnected $\eta_c$ propagators for each ensemble are shown. The parameters $U$, $f$, and $l$
are in lattice units. The total mass shift is $\delta m$ as defined in the text; the first error on
the values is statistical, the second originates from various systematic effects as explained in the last
section. The mass shift due exclusively to the effects of the anomaly is
$\delta m_U$.}
\label{tab:results}
\end{table}

\end{document}